\documentclass[aps,prl,showpacs,twocolumn,superscriptaddress]{revtex4-1}
\usepackage{graphicx}
\usepackage{dcolumn}
\usepackage{bm}
 \usepackage{multirow}

\usepackage[alsoload=hep]{siunitx}

\newcommand{\lisix}[0]{\ensuremath{^6}Li}

\newunit{\atomicunit}{\textmd{a{.}u{.}}}			

\begin{document}

\title{An all-optical atom trap as a target for MOTRIMS-like collision experiments} 

\author{S. Sharma}
\affiliation{Physics Department and LAMOR, Missouri University of Science \& Technology, Rolla, MO 65409, USA}

\author{B.~P. Acharya}
\affiliation{Physics Department and LAMOR, Missouri University of Science \& Technology, Rolla, MO 65409, USA}

\author{A.~H.~N.~C. De Silva}
\affiliation{Physics Department and LAMOR, Missouri University of Science \& Technology, Rolla, MO 65409, USA}

\author{N.~W. Parris}
\affiliation{Physics Department and LAMOR, Missouri University of Science \& Technology, Rolla, MO 65409, USA}

\author{B.~J.  Ramsey}
\affiliation{Physics Department and LAMOR, Missouri University of Science \& Technology, Rolla, MO 65409, USA}

\author{K.~L. Romans}
\affiliation{Physics Department and LAMOR, Missouri University of Science \& Technology, Rolla, MO 65409, USA}

\author{A. Dorn}
\affiliation{Max Planck Institute for Nuclear Physics, Saupfercheckweg 1, 69117 Heidelberg, Germany}

\author{D. Fischer}
\email{fischerda@mst.edu}
\affiliation{Physics Department and LAMOR, Missouri University of Science \& Technology, Rolla, MO 65409, USA}


\begin{abstract}
Momentum-resolved scattering experiments with laser-cooled atomic targets have been performed since almost two decades with MOTRIMS (Magneto-Optical Trap Recoil Ion Momentum Spectroscopy) setups. Compared to experiments with gas-jet targets, MOTRIMS features significantly lower target temperatures allowing for an excellent recoil ion momentum resolution. However, the coincident and momentum-resolved detection of electrons was long rendered impossible due to incompatible magnetic field requirements. Here we report on a novel experimental approach which is based on an all-optical \lisix\ atom trap that -- in contrast to magneto-optical traps -- does not require magnetic field gradients in the trapping region. Atom temperatures of about 2\,mK and number densities up to 10$^9$\,cm$^{-3}$ make this trap ideally suited for momentum-resolved electron-ion coincidence experiments. The overall configuration of the new trap is very similar to conventional magneto-optical traps. It mainly requires small modifications of laser beam geometries and polarization which makes it easily implementable in other existing MOTRIMS experiments.
\end{abstract}

\maketitle

\section{Introduction}

In the last 3 decades, cold target recoil ion momentum spectroscopy (COLTRIMS) \cite{Ullrich2003,Doerner2000} became an indispensable experimental tool in atomic physics revealing unprecedented insights into the correlated dynamics of atoms and their constituents as well as in their interactions with external fields. With this technique it is possible to measure momentum vectors of atomic fragments after ionization events with high resolution down to extremely small kinetic energies. Due to the detailed views into the motion of atomic particles, these spectrometers are also often dubbed 'reaction microscopes' (ReMi). A key ingredient of reaction microscopes is the atomic or molecular target gas which has to be prepared at possibly low temperatures. This is necessary since any statistical thermal motion in the initial state would impair the final momentum resolution in particular for heavy fragments (i.e.\ ions). In most experiments, supersonic gas jets are employed forming a target beam with typical temperatures of few Kelvin. In so-called MOTRIMS experiments, optical cooling is used in magneto-optical traps (MOT) (for a review see \cite{DePaola2008}). Here, substantially lower temperatures (mK or below) can be achieved and target atoms can be prepared in excited and polarized states (e.g.\cite{Zhu09,Leredde2013,Hubele2013}). 

Magneto-optical traps require large magnetic-field gradients in the trap region. While the field affects the momentum measurements of recoil ions only slightly, it renders the momentum-resolved detection of electrons impossible. This apparent incompatibility of magneto-optical trapping and electron momentum spectroscopy has recently been resolved in a MOTReMi experiment \cite{Fischer2012}. Here, a quadrupole magnetic field is switched periodically on and off, thereby alternating between magneto-optical trapping and data acquisition cycles, respectively.  Despite the success of this approach, the experimental challenges should not be underestimated. The switching of the magnetic field substantially complicates the design of the experiment, its operation, as well as the analysis of the data acquired with the setup. Eddy currents have to be considered, the data acquisition and fast field switching need to be synchronized, and temporal fluctuations induced by the field switching must be accounted for in the data analysis. 
 
In this paper we report on an alternative optical cooling and trapping scheme which we implemented in the MOTReMi experiment. Similar to MOTs, the present scheme relies on the interaction of atoms with near resonant light, but in the present all-optical configuration no magnetic field gradient is required. Although the present trapping scheme has resemblances to earlier realized configurations \cite{Chu1987,Walker92,Walker92a,Hope94}, it stands out since the trap can be operated with homogeneous magnetic fields of more than 10 Gauss. This makes it an ideal target for collision experiments in reaction microscopes, which could easily be implemented in other existing MOTRIMS experiments.

\section{Experiment}

This experiment was performed with the MOTReMi, which is a unique combination of a magneto-optical trap for target preparation and a fully equipped reaction microscope for the momentum resolved detection of electrons and recoil ions. Here, only the components relevant for the new trapping mechanism are briefly discussed. A more detailed description of the momentum spectrometer and its overall design can be found in \cite{Hubele2015}.

\begin{figure}
\centering
\includegraphics[width=70mm]{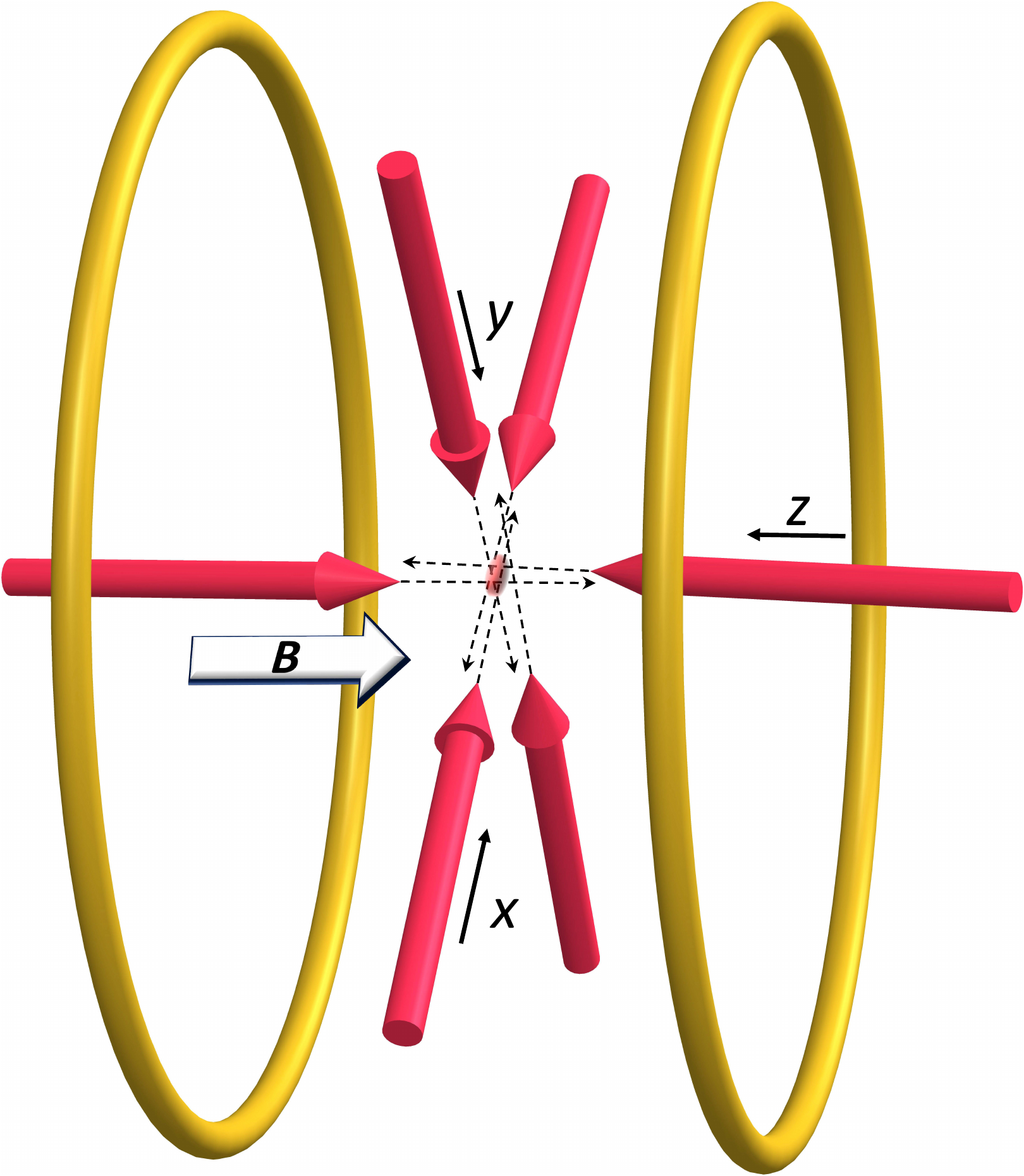}
\caption{Schematic view of the experimental setup. The retro-reflected laser beam pairs are slightly offset to one another. The magnetic field is collinear to the $z$-axis and generated by a pair of Helmholtz coils outside the vacuum chamber.}
\label{fig:setup} 
\end{figure}

The atom trap is located in a vacuum chamber with a background pressure of about 10$^{-10}$\,mbar and loaded with a cold beam of \lisix\ atoms from a 2-dimensional MOT. As in conventional magneto-optical traps, the atoms are cooled and trapped by three retro-reflected pairs of laser beams aligned along mutually (nearly) orthogonal axes. We choose the following coordinate system (see figure \ref{fig:setup}): Two laser beam pairs are oriented along the x- and y-direction, respectively. Due to the momentum spectrometer, the third pair is not precisely perpendicular to the x- and y-axis because particle detectors obstruct the optical access to the trap volume along the z-axis (for details see \cite{Hubele2015}). Therefore, this beam has a small angle of 12.5$^\circ$ with respect to the z-axis and points along the vector ($-\sqrt{2} \text{sin}12.5^\circ, -\sqrt{2} \text{sin}12.5^\circ,  \text{cos}12.5^\circ$). For the MOT operation of the atom trap, there is a pair of in-vacuum anti-Helmholtz coils which creates a quadrupole magnetic field in the trap region. However, for the trapping mode discussed here, these coils were not in operation. Additionally, a homogeneous magnetic field in the z-direction up to 15 Gauss can be generated with a large pair of Helmholtz coils located outside of the vacuum chamber. 

The laser beams are provided by a tapered-amplifier diode laser system whose frequency is stabilized slightly below the \lisix\ D2-transition frequency from the $^2$S$_{1/2}$ ground state to the $^2$P$_{3/2}$ excited state ($\lambda = 671$\, nm). The ground state of \lisix\ has a hyperfine splitting of about 230 MHz. In order to avoid optical pumping to a dark state, the cooling beams need to contain two frequencies that excite transitions from both hyperfine levels. This is achieved by using an electro-optical phase-modulator (EOM) producing sidebands shifted by $\pm 230$\,MHz with respect to the cooler frequency (corresponding to the transition $^2$S$_{1/2}\to ^2$P$_{3/2}$ with the total atomic angular momenta F=3/2 and F=5/2, respectively). In the present experiment, the beams had diameters of about 10-15\,mm with total powers of 15 to 25\,mW. About 50\,\% of the power is at the cooler frequency and 25\,\% is up-shifted by 230\,MHz to the re-pumper frequency ($^2$S$_{1/2}$, F=1/2 $\to$ $^2$P$_{3/2}$, F=3/2). Another 25\,\% is shifted 230\,MHz below the cooler frequency and not in resonance to either transition.

Compared to magneto-optical traps, the present trapping scheme is much more sensitive to experimental parameters like beam positions, polarization, intensities, and frequency detuning. Similar to earlier reported configurations like the `supermolasses' \cite{Chu1987} or the vortex trap \cite{Walker92}, the positions of any two counter-propagating laser beams are slightly misaligned. Consequently, the adjustment of the beam positions and other parameters is not straightforward but is a tedious procedure. The general strategy to achieve stable trapping will be described in the following.

First, the atoms are trapped in a conventional MOT configuration. In this scheme, a quadrupole magnetic field is generated around the trap position with the anti-Helmholtz coils, and all laser beams are circularly polarized by means of $\lambda$/4 wave-plates with any two retro-reflected beams having opposite angular momentum ($\sigma^+-\sigma^-$ configuration). Next, a homogeneous magnetic field in the z-direction is superposed and stepwise increased up to about 7\,Gauss.  Due to this field, the equilibrium position of the trap (i.e.\ of the local minimum of the magnetic field strength) shifts and the atom number drops. By adjusting positions, polarization, detuning of the beams, and the parameters of the atom source (i.e.\ the 2D MOT), the atoms' signal can (at least partially) be retrieved. Note, that the trap position should stay unaltered compared to the initial MOT and should not shift with the zero magnetic field position. At sufficiently high magnetic fields, the location of the atom cloud can only be retained if the polarization of the two laser beams along the z-axis is identical because the Zeeman Effect shifts the $\sigma^+$ transition to far from resonance which results in an imbalance of forces for the $\sigma^+-\sigma^-$ configuration. Therefore, the $\lambda$/4 wave-plate, that flips the helicity of the retro-reflected z-beam from $\sigma^-$ to $\sigma^+$, has to be removed, which again makes a readjustment necessary. The configuration achieved in this way, resembles closely the spontaneous-force atom trap described by \citet{Walker92a}.

Thereafter, the quadrupole magnetic field is stepwise reduced and in each step all experimental parameters are again adjusted iteratively. When the quadrupole magnetic field is entirely switched off, the $\lambda$/4 wave-plates used for the beams in x- and y-direction can be removed and replaced by $\lambda$/2 wave-plate for each incoming beam. The trap was operated with two different laser polarization schemes. In the first one, referred to as $\sigma$-configuration, the light was circularly $\sigma^-$-polarized for the beams longitudinal to the external magnetic field and linearly polarized for the transverse ones. For the latter the electric field vectors were perpendicular to the B field allowing only for the excitation of $\sigma$-transitions. In the second scheme, in the following referred to as $\pi$-configuration, all laser beams were linearly polarized with the polarization vectors of the transverse beams being parallel to the external magnetic field driving only $\pi$-transitions.

Although stable trapping has been achieved for a large variety of experimental parameters, each configuration features a very narrow optimum, meaning a tiny variation of only one parameter can already result in the loss of the trapped atoms. All configurations had in common, that the laser beams were misaligned relative to the trap center resulting in a vortex force. An example of laser beam parameters is listed in table \ref{tab:beams}.

 \begin{table}
 \caption{Example set of  laser beam parameters.}
  \centering
 \scalebox{0.92}{
 \begin{tabular}{|l|c|c|c|c|}
 \hline Coordinate axis & X & Y & Z' \\ 
 
  \hline \hline
&&&\\
Beam cross sections
 & \multirow{3}{*}{\includegraphics[ height=18mm] {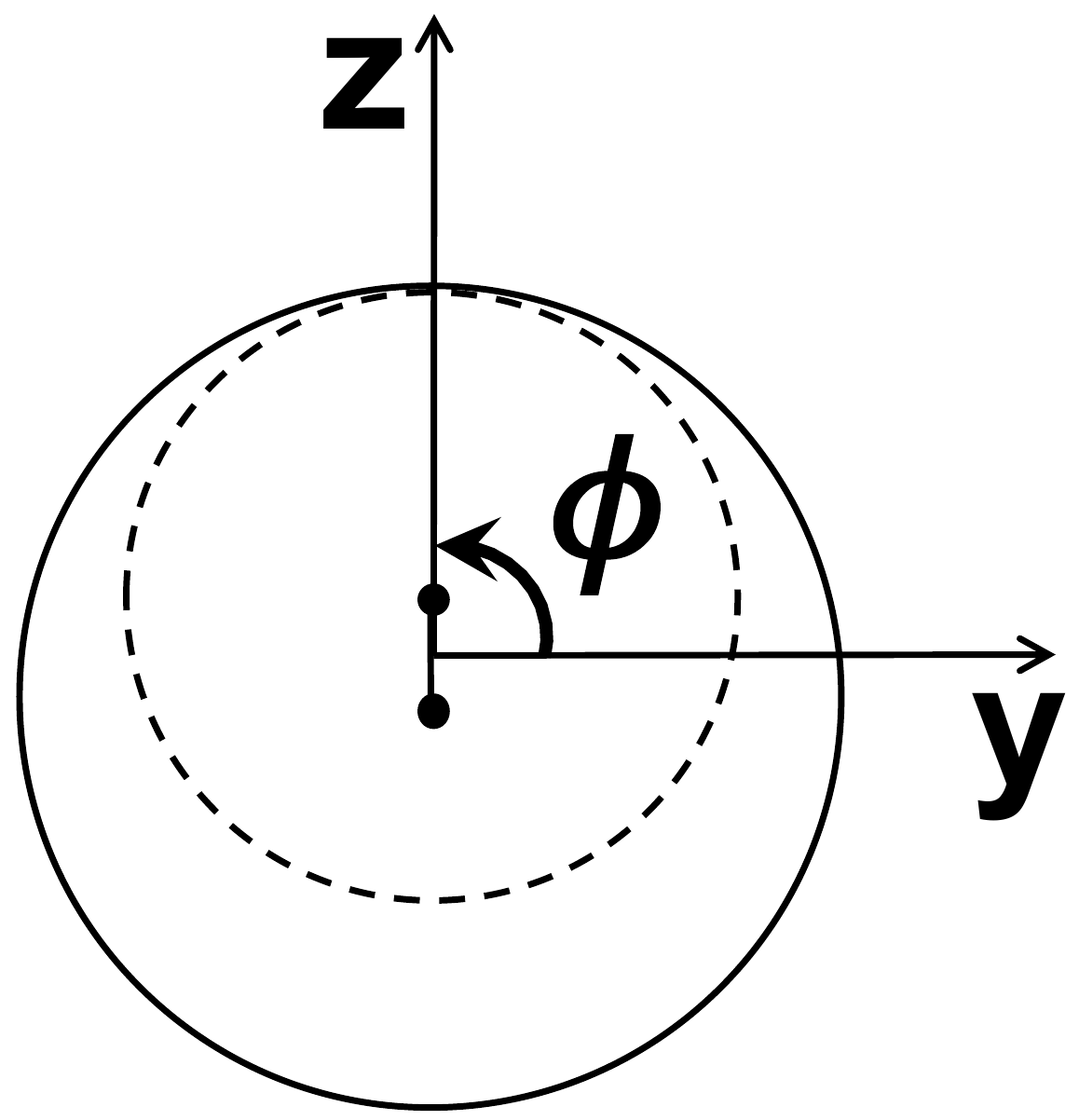}} & \multirow{3}{*}{\includegraphics[height=18mm] {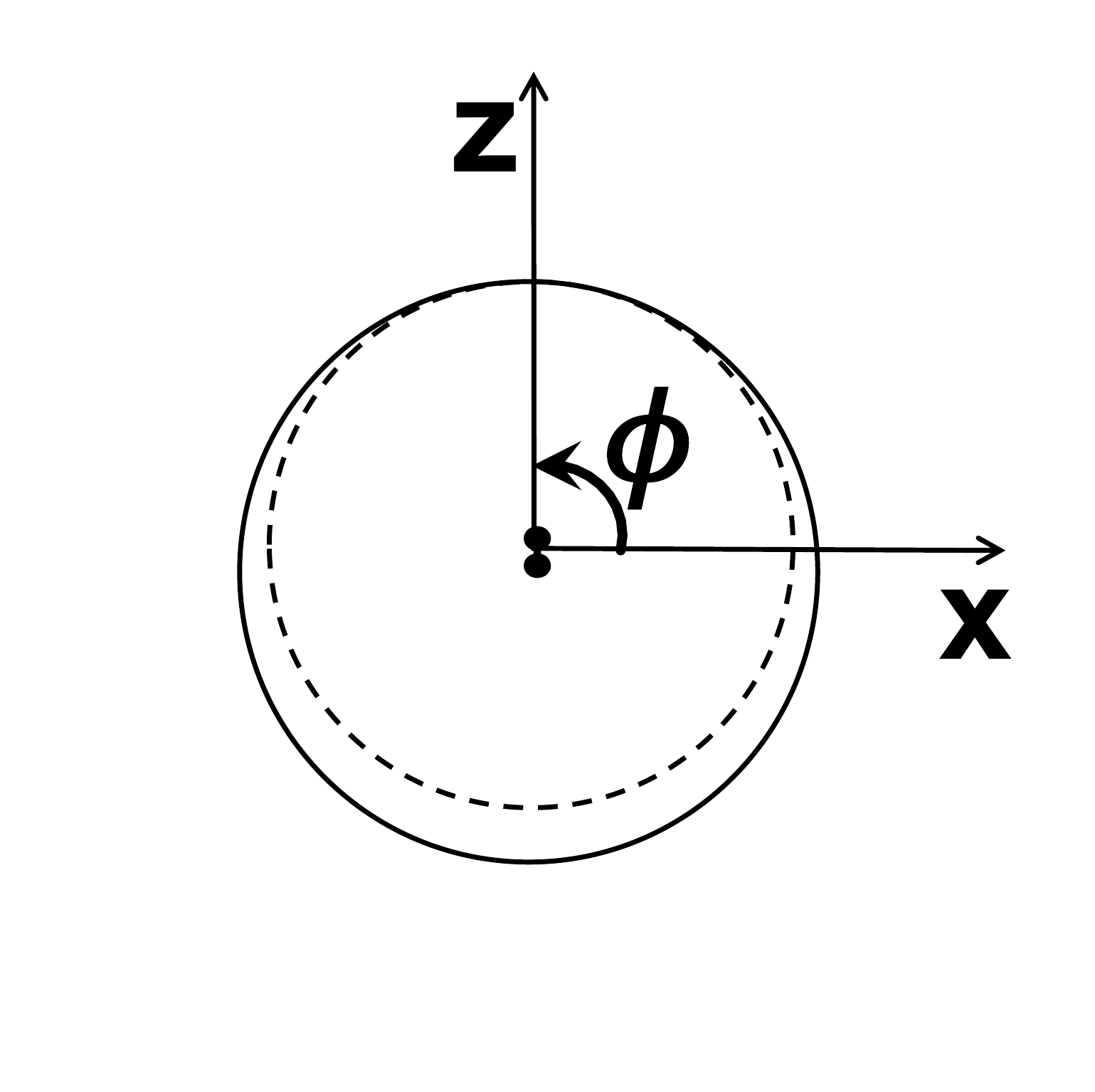}} & \multirow{3}{*}{\includegraphics[ height=18mm] {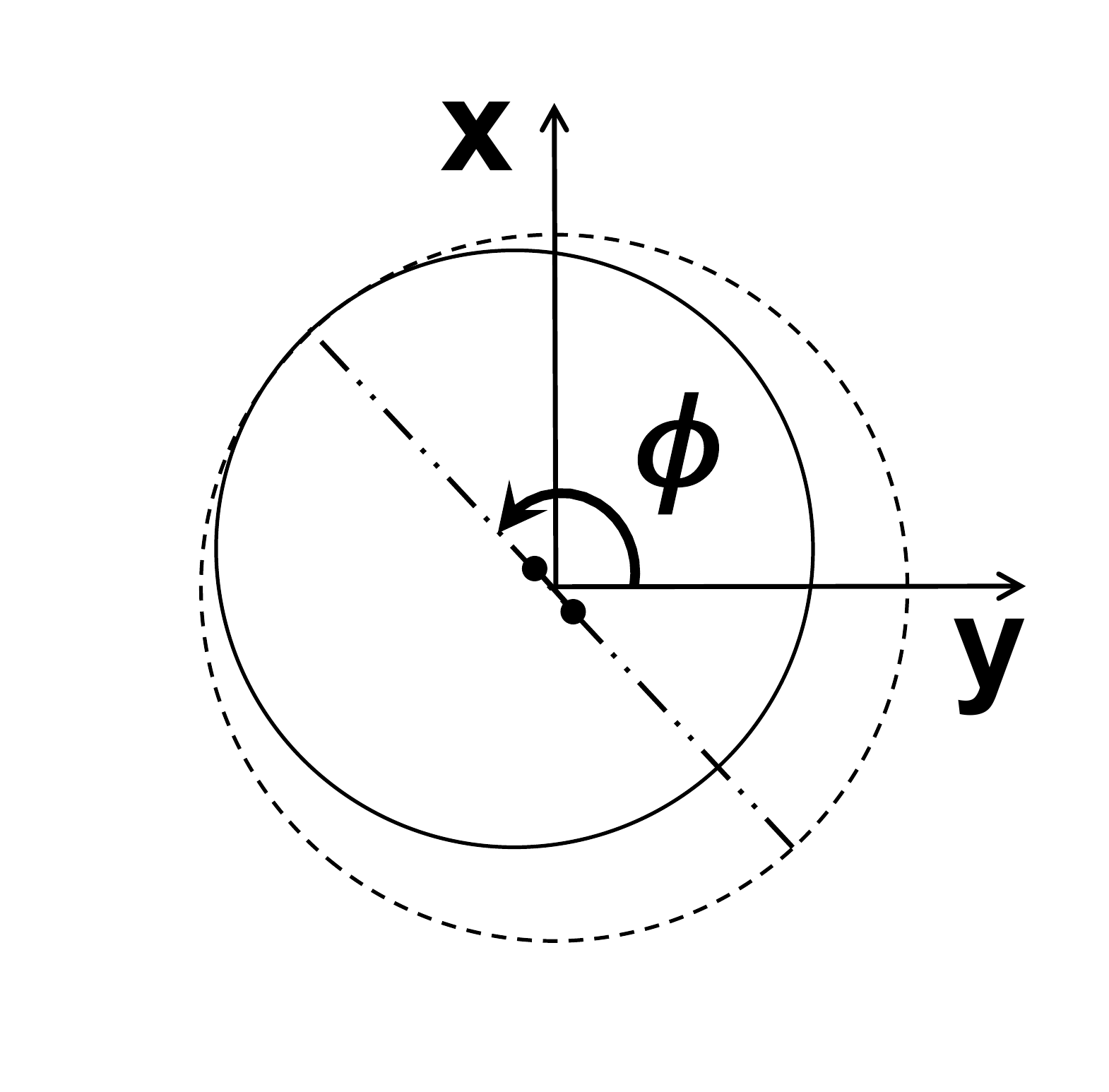}}\\ 
 \quad ingoing (solid) &&&\\
 \quad retro-refl.\ (dashed) &&&\\
 &&&\\
 &&&\\
 &&&\\
Beam diameters (mm) & 11.5/8.5 & 10.6/9.6 & 11/13\\
(ingoing/retro-refl.) &&&\\
 &&&\\
Power (mW) & 24/21.6 & 16/14.4 & 18/16.2 \\
(ingoing/retro-refl.) &&&\\
&&&\\
 Displacement & \SI{1.5}{\milli\meter} & \SI{0.5}{\milli\meter} & \SI{1}{\milli\meter} \\
  (between beam centers) &&&\\
 &&&\\
 Angle $\phi$ of coaxal / & $\sim$ 90$^\circ$/Y & $\sim$ 90$^\circ$/X & $\sim$ 135$^\circ$/Y \\
 reference axis & & &\\
 &&&\\
  \hline 
  \end{tabular}}
 \label{tab:beams}
 \end{table}

\section{Characterization of the laser-cooled gas cloud}
The feasibility of highly differential atomic collision experiments depends strongly on target parameters such as the target density and temperature because these attributes influence the achievable coincidence rate and resolution, respectively. However, knowing the target cloud properties is not only important for the design and analysis of scattering experiments, it also allows conclusions to be drawn on the trapping mechanism. 

To characterize the \lisix\ cloud, we determined the trapped atom number, the number density, the characteristic loss and loading rates, the cloud temperature, and the degree of polarization of the atoms in the excited 2 $^2$P$_{3/2}$ state. For most measurements fluorescence imaging was employed using three CMOS cameras positioned at relatively small angles (10$^\circ$ to 20$^\circ$) with respect to the x-, y-, and z-axis, respectively. For the measurement of the atom number we used additionally absorption imaging and obtained a consistent result. All the measured properties are listed in table \ref{tab:trapprop}. Some of the measurements with the $\sigma$-configuration are discussed in detail below.

\begin{table}
 \caption{Typical properties of the gas cloud.}
 \centering
 \begin{tabular}{|l|c|c|}
 \hline & $\sigma$-config. & $\pi$-config. \\
  \hline \hline
Lifetime (1/$\mathit{\Gamma}$)  & $\sim$ 5\,s &  $\sim$ 4\,s\\
No.\ of trapped atoms ($N_\text{eq}$)  &$\sim$ $10^7$ & $\sim$ $10^7$\\ 
No.\ density  &$\sim$ $10^9$/cm$^3$ & $\sim$ $10^9$/cm$^3$\\ 
Temperature: & &\\
\ \ \ \  X-axis & 2.5\,mK & 2\,mK\\
\ \ \ \ Y-axis & 700\,$\mu$K  & 2\,mK\\
\ \ \ \ Z-axis & 2\,mK & $>5$\,mK\\
Polarization of fluorescence: & &\\
\ \ \ \ $\sigma^- $ & 93\,\% & 43\,\% \\
\ \ \ \ $\pi $ & 5\,\% & 43\,\% \\
\ \ \ \ $\sigma^+ $ & 2\,\% & 14\,\% \\
Degree of polarization & 90\,\% & 29\,\% \\
\hline 

  \end{tabular}
 \label{tab:trapprop}
 \end{table}

\subsubsection*{Atom number and density}

With the configuration discussed above, trapped atom numbers of about 10$^7$ are observed corresponding to number densities of $10^9$\,cm$^{-3}$ in a cloud of 1 to 2\,mm diameter. While such a target density is high enough for the study of interactions with high-intensity charged-particle or photon beams in collision experiments, it is still substantially lower than the densities in other magneto-optical traps which -- for heavier atomic species -- can be up to 4 orders of magnitude higher. In order to identify the factors limiting the maximum atom number in the present configuration, the loss mechanisms and the loading rate of the \lisix\ atoms were studied.

Generally, the trapped atom number $N$ follows the simple rate equation

\begin{equation}
\frac{\text{d}N}{\text{d}t}=L-\mathit{\Gamma} N-\beta \int n(\vec{r})^2 \text{d}^3r
\label{equ:rate}
\end{equation}

with the loading rate $L$, the linear loss factor $\mathit{\Gamma}$ due to collisions with the residual gas or due to the heating caused by photon emission, and the two-atom loss coefficient $\beta$ due to mutual collisions between two lithium atoms. For not too high densities, the two-atom loss term can be approximated by $\beta' N^2$. This approximation is valid if the shape of the density distribution $n(\vec{r})$ does not change with the total atom number $N$ (in MOTs this condition is typically fullfilled for densities below about 10$^{10}$\,cm$^{-3}$ \cite{Overstreet2005}). In this case, the loading rate $L$, the linear loss factor $\mathit{\Gamma}$, and the effective two-atom loss factor $\beta'$ can easily be measured switching the atom beam from the 2D-MOT on and off thereby alternating the loading rate $L$ between zero and its maximum value.

The general solution of the differential equation \ref{equ:rate} can be expressed in the form
\begin{equation}
N(t)=N_\text{eq}\frac{1-\nu\xi e^{-\gamma t}}{1+\xi e^{-\gamma t}}
\label{equ:soln}
\end{equation}
with the equilibrium number of atoms $N_\text{eq}=N(\infty)$ being
\begin{equation}
N_\text{eq}=\frac{\sqrt{4\beta' L+\mathit{\Gamma}^2}-\mathit{\Gamma}}{2\beta'}
\label{equ:nmax}
\end{equation}
and with the coefficients $\gamma=2\beta' N_\text{eq}+\mathit{\Gamma}$ and $\nu=L/(\beta' N_\text{eq}^2)$ as well as $\xi=(N_\text{eq}-N(0))/(\nu N_\text{eq}+N(0))$. 
For the following discussion, two situations are of special interest. First, the drop-off of the atom number initially being $N(0)=N_0$ after setting $L\to 0$. 
\begin{equation}
N(t)=\frac{N_0 e^{-\mathit{\Gamma} t}}{1+N_0 \beta'/\mathit{\Gamma}-N_0 \beta' e^{-\mathit{\Gamma} t}/\mathit{\Gamma}}
\label{equ:decay}
\end{equation}
The second scenario is the loading of the initially empty trap, i.e. $N(0)=0$ and $L>0$, where the atom number follows the curve
\begin{equation}
N(t)=N_\text{eq}\frac{1-e^{-\gamma t}}{1+e^{-\gamma t}/\nu}
\label{equ:loading}
\end{equation}
%
\begin{figure}
\centering
\includegraphics[width=80mm]{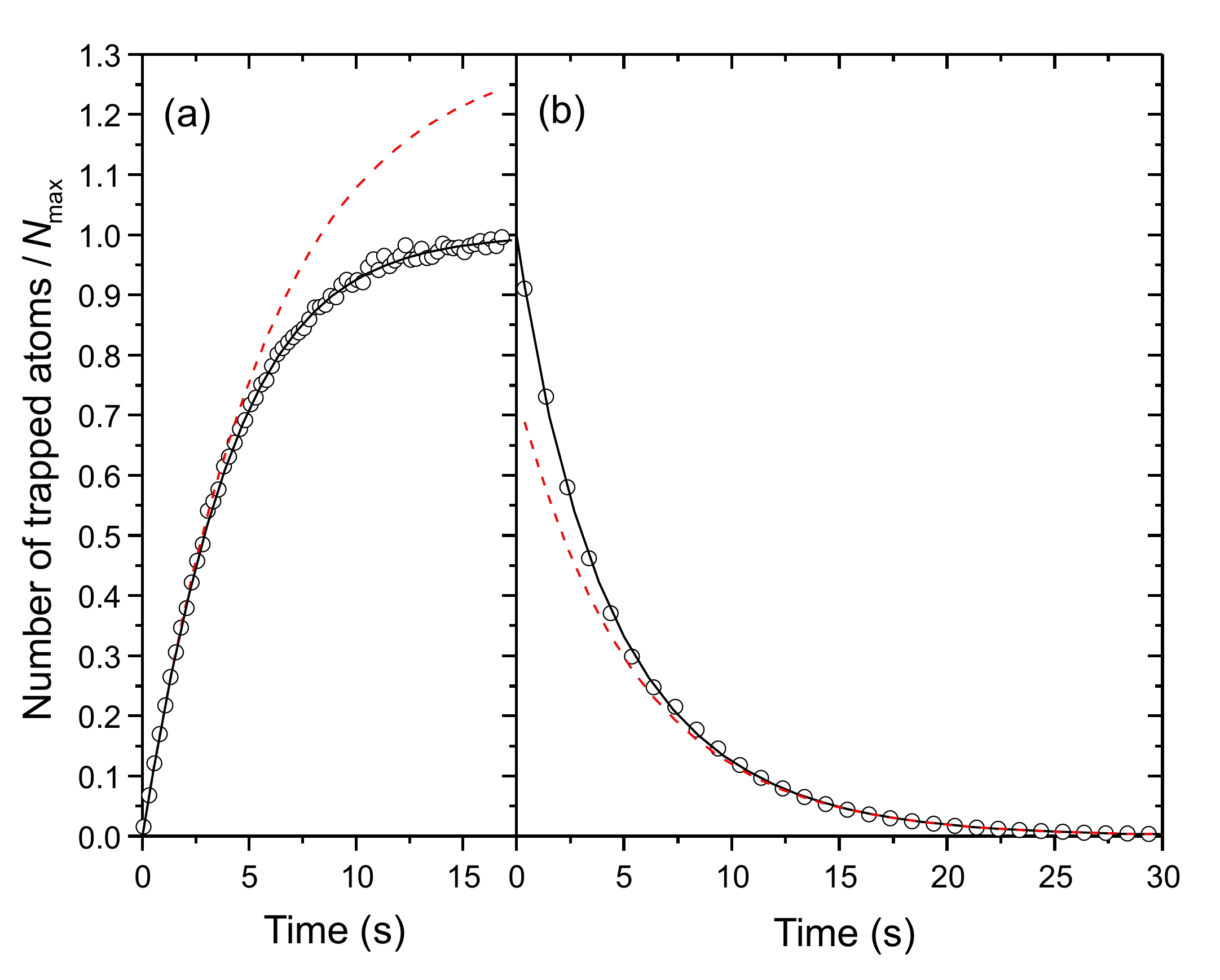}
\caption{Loading (a) and depletion (b) curve of the trap. Experimental data are shown as open circles, the solid lines correspond to the fits according to equation \ref{equ:decay} and \ref{equ:loading}, respectively. The dashed lines are exponential curves fitting the data for low atom number densities (i.e.\ where the two-atom loss term is negligible).}
\label{fig:loaddec} 
\end{figure}

In figure \ref{fig:loaddec} (a) and (b), the atom trap population is shown as a function of time for the loading and the decay of the trap, respectively. The fits using Eq.~\ref{equ:decay} and \ref{equ:loading} show excellent agreement with the experimental data, while a pure exponential fit (i.e.\ assuming $\beta=0$) describes the data only for very low atom numbers. Both the loading and the decay curve contain information on the $\mathit{\Gamma}$ as well as the $\beta'$ parameter and there is reasonable agreement between the two independent fitting results. According to the fits, the loading rate is about $L=2\cdot10^6$/s, the linear decay rate is about $\mathit{\Gamma}=0.17$/s, and the effective two-atom loss rate is roughly $\beta'=6\cdot 10^{-9}$/s, meaning that for a fully loaded trap about every fourth atom is lost due to mutual \lisix\ collisions. Assuming a three-dimensional Gaussian distribution for the atom cloud density, the two-body loss coefficient can be calculated from $\beta'$ and it is $\beta\approx 8\cdot 10^{-11}$\,cm$^3$/s.

Compared to standard magneto-optical lithium traps (e.g.\ \cite{Ridinger11}), the present trap features $\mathit{\Gamma}$ and $\beta$ parameters which are only slightly higher indicating that the effective trap potential is shallower than those of conventional MOTs. A substantially larger discrepancy is observed for the loading rate $L$ which is in the present study about 3 orders of magnitude lower than reported for earlier experiments \cite{Ridinger11,Tiecke09}. This can partially be explained with the shallower trapping potential, but it has mainly to be attributed to the low atom flux of the loading beam. The design of the present atom source is very similar to the 2D-MOT described in \cite{Tiecke09}, however, the available laser beam power for the trapping and pre-cooling of the atoms is by a factor of 5 lower resulting in a substantially reduced cold atom flux. According to Eq.\ \ref{equ:nmax}, enhancing the loading rate to $L=10^9$/s would lead to an increase of number density to almost 10$^{11}$\,cm$^{-3}$ (neglecting the effect of geometrically increasing the trap size for high number densities \cite{Overstreet2005}). 

Furthermore, the relation between the steady state atom number $N_\text{eq}$ and the cooling laser beam intensities was studied. Notably, $N_\text{eq}$ did not saturate in the investigated regime but rather increased monotonically with raising the laser intensities. This is remarkable, because  the saturation intensity $I_\text{sat}$ (2.54\,W/cm$^2$ for the lithium D2-transitions) is well below the present intensities, which are on average about 2.5 to 7.5 times $I_\text{sat}$ for each individual beam. In this domain, the spontaneous force saturates and in a classical optical molasses description \cite{Lett89} no significant dependence of the atom number on the laser intensity is expected. As will be discussed in detail below, this is clear evidence that the present trapping mechanism cannot solely be explained by the resonant scattering of photons.

\subsubsection*{Cloud temperature}
\begin{figure}
\centering
\includegraphics[width=70mm]{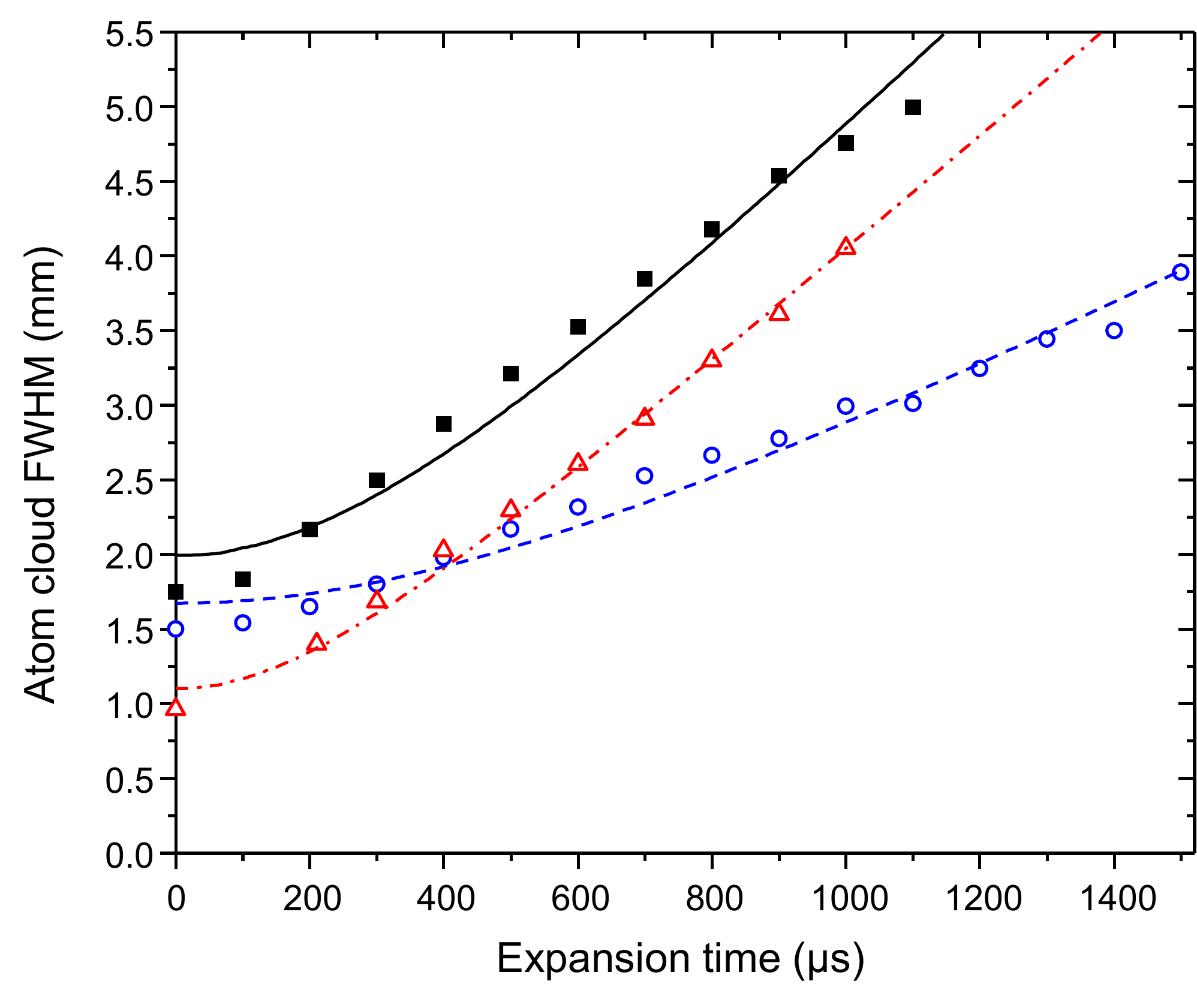}
\caption{Thermal expansion of the atom cloud. The cloud width is plotted as a function of the expansion time. The symbols represent the experimental data for the x (squares), y (open circles), and z-direction (open triangles). The solid, dashed, and dash-dotted lines depict the respective fitting curves.}
\label{fig:thermexp} 
\end{figure}

The temperature of the atoms was determined by measuring the ballistic expansion of the cloud when no forces are exerted on the atoms. To this end, the cooling lasers were switched off for short periods $t$ and the width of the cloud was determined after the switch-off by fluorescence imaging. In figure \ref{fig:thermexp}, the cloud widths in x, y, and z directions are plotted as a function of the time $t$. For a gas initially in thermal equilibrium (more precisely, the initial 6-dimensional phase-space distribution of the gas being represented by a simple product of six Gaussians, one for each component), the spatial distribution along each axis for any given time $t$ should be well reproduced by the convolution of two Gaussians, the first representing the initial distribution and the second the thermal expansion. The corresponding fitting curves are shown as lines in figure \ref{fig:thermexp}. From the fit, the thermal velocity and, thus, the temperature of the gas can be extracted.

There are two notable observations made in the temperature measurements: First, the temperature is not identical along the three coordinate axes. The analyses yield about 2\,mK in the x and z-directions and about 700\,$\mu$K in the y-direction. These temperatures are substantially larger than the Doppler temperature (140\,$\mu$K for \lisix ). 
Second, the fitting model used has a relatively poor agreement with the experimental data for the first 200 to 300\,$\mu$s after switching the lasers off. Both observations can be explained by a vortex motion of the atoms due to the misalignment of the laser beams. This motion results in higher velocities in the plane of rotation leading to higher temperatures in the x-z plane. Moreover, the velocity distribution becomes dependent on the atoms' position resulting in a more complex time dependence than is expected with our fitting model.

\subsubsection*{Atomic polarization}
\lisix\ features a complex multi-level structure with two hyperfine levels in the $^2$S$_{1/2}$ ground state (F=1/2 and 3/2) and three hyperfine levels in the excited $^2$P$_{3/2}$ state (F=1/2, 3/2, and 5/2). In contrast to heavier alkali-metal atoms such as rubidium or cesium, the energy separation of the excited hyperfine levels in lithium is too small to be resolved by the cooling laser beams. This makes it generally more challenging to disentangle the distribution of populated states. Nevertheless, there are at least three ways to obtain information on the involved transitions and the atomic orientation: First, the polarization of the cooling laser beams; second, the shift of the optimum cooling laser frequency with the strength of the external magnetic field; and third, the polarization of the emitted fluorescence light.

\begin{figure}
\centering
\includegraphics[width=80mm]{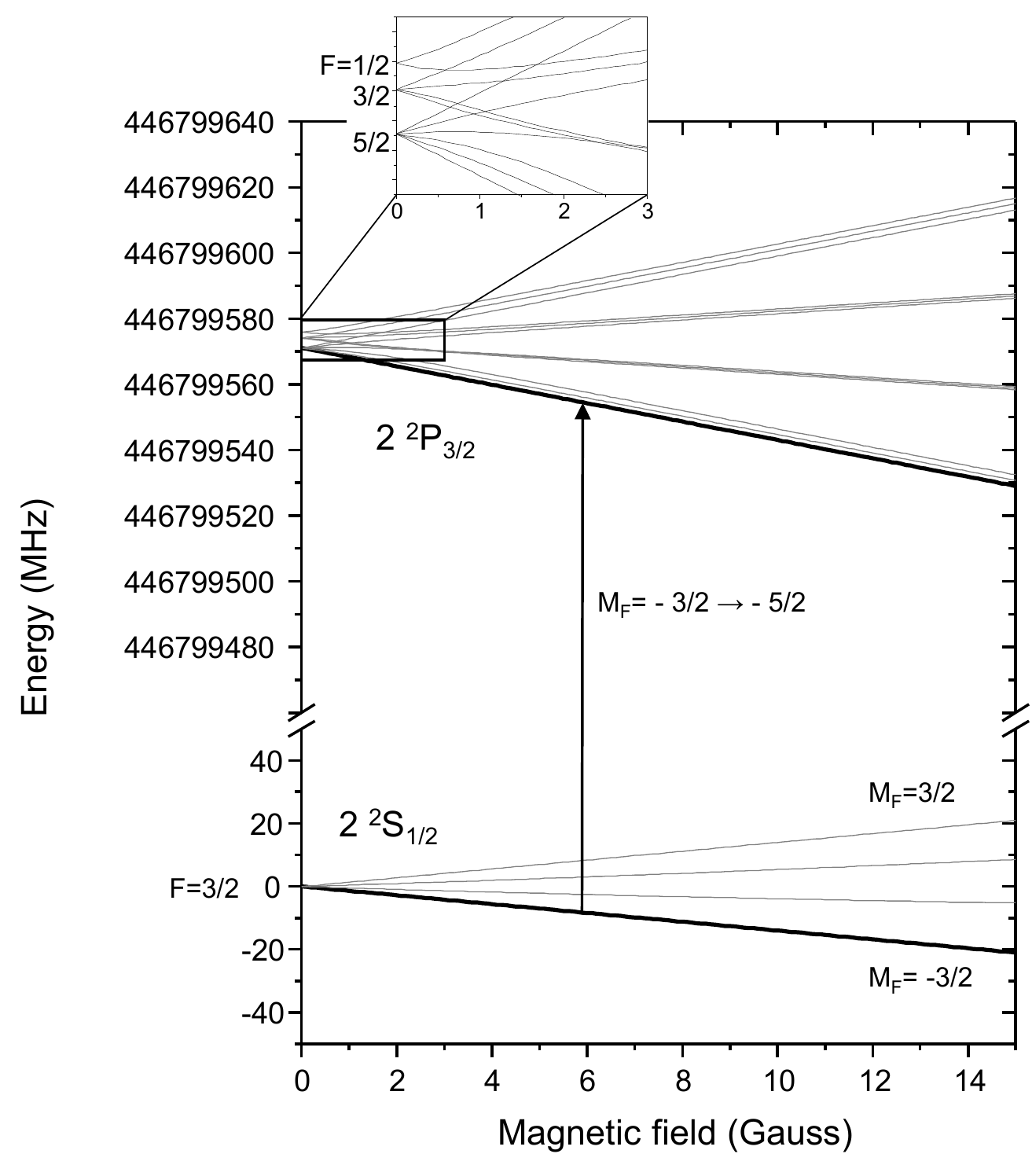}
\caption{Zeeman splitting of the  $^2$S$_{1/2}$ (F=3/2) and the $^2$P$_{3/2}$ levels of \lisix\ (from \cite{Gehm03}).}
\label{fig:zeeman} 
\end{figure}

\begin{figure}
\centering
\includegraphics[width=70mm]{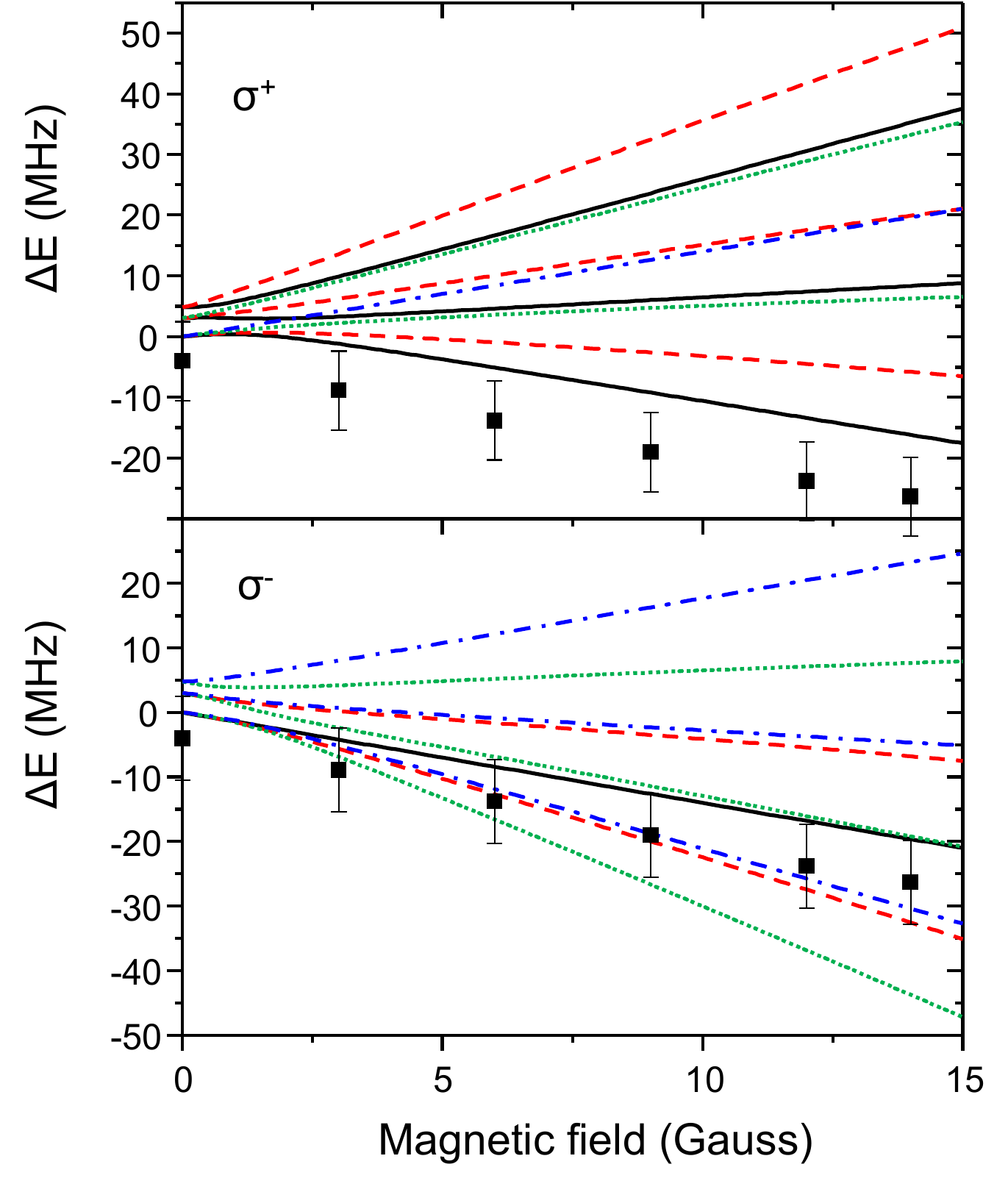}
\caption{Energy shift of $\sigma^+$ (top) and $\sigma^-$-transitions (bottom) as a function of the external magnetic field $B$. Zero energy shift corresponds to $^2$S$_{1/2}$ (F=3/2)$\to^2$P$_{3/2}$ (F=5/2) transitions at $B=0$. Different line types are for the ground state magnetic sub-levels M$_\text{F}$=-3/2 (solid), -1/2 (dashed), 1/2 (dotted), and 3/2 (dash-dotted). The data points correspond to the experimental laser frequency at optimum trapping efficiency (see text).} 
\label{fig:sigma+-} 
\end{figure}

In the present $\sigma$-configuration, the laser beams oriented along the z-direction are circularly polarized driving only $\sigma^-$ transitions. The other laser beams are linearly polarized with their electric field vectors being perpendicular to the z-axis allowing for both $\sigma^+$ and $\sigma^-$ (but not $\pi$) transitions. The energy levels of the involved magnetic sub-states are shown in figure \ref{fig:zeeman} as a function of the magnetic field strength. The corresponding transition energies are displayed in figure \ref{fig:sigma+-} (a) and (b)  for  $\sigma^-$ and $\sigma^+$ transitions, respectively. The data points in the graph correspond to the incoming laser beam frequency with the error bars accounting for the natural linewidth (5.8\,MHz), the estimated Doppler broadening (for 2\,mK), and the bandwidth of the laser system ($\sim 1$\,MHz).

As seen from Fig.\ \ref{fig:sigma+-}, there are $\sigma^-$ transitions close to the cooling laser frequency for all ground state sub-levels, while for the  $\sigma^+$ transitions the light is significantly farther off resonance except for the M$_\text{F}=-3/2$ ground state. Notably, the electric dipole matrix element of the corresponding  $\sigma^+$ transition (M$_\text{F}=-3/2$ to $-1/2$) is at least a factor of 2.4 (depending on the magnetic field strength) smaller than of the $\sigma^-$ transition from the same ground state. Therefore, $\sigma^+$ transitions are generally suppressed resulting in significant optical pumping to the states with the smallest M$_\text{F}$. This conjecture is underpinned by the dependence of the incoming laser frequency on the magnetic field which follows closely the Zeeman shift of the transition  M$_\text{F}=-3/2 \rightarrow -5/2$ and is consistently about 4 to 6\,MHz red-detuned to this resonance.

Information on the polarization of the target cloud can also be obtained by measuring the polarization of the fluorescence signal \cite{Hubele2015}. In the present configuration, the fluorescence photons are emitted in $\sigma^-$, $\pi$, and $\sigma^+$ transitions with relative contributions of 93\,\%, 5\,\%, and 2\,\% corresponding to a degree of polarization of 90\,\%. This high degree of polarization is consistent with the optical pumping mechanism discussed above resulting in a closed cooling cycle between the $(\text{F},\text{M}_\text{F})=(3/2,-3,2)$ and $(5/2,-5/2)$ states (indicated by the arrow in Fig.\ \ref{fig:zeeman}) with only a small leak due to  $\sigma^+$ excitation.

For the $\pi$-configuration a much lower degree of polarization is observed (see table \ref{tab:trapprop}). Here, all cooling beams are linearly polarized with their electric field vectors being parallel to the magnetic field for the beams in x and y-directions and, hence, driving only $\pi$-transitions. The beams along the z-axis can lead to $\sigma^+$ and $\sigma^-$ transitions with the latter once being dominant due to a smaller detuning. Consequently, optical pumping effects are inhibited in this configuration.

\section{Trapping mechanism}

In general, there are two types of forces acting on a two-level atom exposed to the field of a monochromatic laser beam: the gradient force (or optical dipole force) oriented along the field intensity gradient, and the scattering force (or spontaneous force) in the direction of the Poynting vector. While these forces are very well understood, the explanation of the properties of laser-cooled samples is still extremely challenging (e.g.\ \cite{Prudnikov2015}) owing to the complexity of real-world experimental systems involving e.g.\ atomic multi-level structure, interference-induced intensity patterns, and nonmonochromaticity of the laser beams. These features can give rise to effects such as Sysiphus cooling or the Raman coupling of different hyperfine states. They can also result in ``rectified'' dipole forces providing confinement of atoms. 
In spite of the tremendous successes achieved in the theoretical description of many subtle cooling and trapping mechanisms, there are still questions left unanswered. For instance, there is to our knowledge no complete and consistent explanation for the phenomenon of ``supermolasses'' -- first observed by \citet{Chu1987} -- reported in literature. The present configuration has obvious resemblances to the supermolasses and other atom traps with similar laser beam geometries \cite{Walker92,Walker92a,Hope94}. We too presently do not have a full model that explains all features of our trapped gas cloud. However, based on our experimental results we can still draw conclusions about the effects that are relevant for the dynamics in the present atom trap.

Compared to earlier traps, the present $\sigma$-configuration is particularly clean and simple in two respects: First, any spatial dependence of the forces can only be related to variations in the optical field itself because all other fields (in particular the magnetic field) are homogeneous. Second, position and velocity dependent optical pumping effects, that can give rise to spatial confinement (e.g.\ \cite{Pritchard86, Cooper2013}), are minimized. This is because all laser beams drive the same $\sigma^-$-transitions with respect to a `universal' quantization axis (except for a small contribution of $\sigma^+$-transitions). Therefore, the cooling cycle is almost closed between two magnetic sub-states making it an effective two-level system.

For the present experimental conditions, the force exerted on the atoms  by a single monochromatic laser beam is vastly dominated by the spontaneous force which exceeds the optical dipole force by about six orders of magnitude. This suggests, that the present configuration could be described by a simplistic optical molasses model including only the spontaneous force exerted on two-level atoms. Although it has been shown that a scattering force alone -- if merely proportional to the photon flux -- cannot result in a stable trapping of atoms \cite{Ashkin83} it was earlier claimed that stable trajectories can exist in systems with misaligned laser beams due to a damped vortex motion \cite{Walker92} or due to a dynamical stabilization process similar to the Kapitza pendulum \cite{Bagnato94}. We performed a three-dimensional Monte Carlo simulation using the optical molasses picture described in \cite{Lett89}, where the experimental geometry was accounted for by implementing six Gaussian laser beams with adjustable powers, widths, and positions. We tested this model by examining two scenarios. First, we simulated the trap loss rate due to the random walk of the atoms. Second, we tried to reproduce the trap loading by the atom capture out of the cold beam from the 2D-MOT.





In the first simulation, we considered an ideal optical molasses with perfectly aligned laser beams of equal intensity. Here, the trapping time is limited due to the heating caused by the re-emission of photons and it amounts to about 0.5\,s. These results are consistent with earlier simulations and experimental observations (e.g.\ \cite{Chu1987}). In our experiment there is a substantial imbalance in laser intensities due to losses at the viewports and the back-reflecting mirrors. For laser beam parameters as listed in table \ref{tab:beams} the simulation yields a trapping time of about 30\,ms  which vastly underestimates the experimentally observed trapping time of about 5\,s. 

In order to investigate the capture process of atoms out of the loading beam, we first measured the velocity distribution of the incoming atoms with the MOTReMi by photo-ionizing the \lisix\ atoms from the trap and from the loading beam, respectively, with an ultra-violet laser. By comparing the fragments' momentum distribution, the atom beam velocity was determined to be about 35$\pm 10$\,m/s. We simulated the interaction of the cooling lasers with the atom beam entering the trap region roughly along the bisecting line between the x and the y-axis. It is found that none of the atoms are captured in the trap but they are either back-reflected or guided around the trapping volume without resulting in a significant accumulation of atoms in the trap region.

Both tests of our Monte Carlo model clearly indicate that the classical optical molasses picture is insufficient to describe the dynamics in the present experiment. This is not surprising because many properties of the system have been neglected, among them the multi-level structure of the \lisix\ atoms, the multi-chromaticity of the incoming light, and spatial interference structures of the six laser beams. Moreover, important mechanisms like the AC Stark shift and the stimulated emission of photons are not considered in the model even though they are expected to be significant at the present intensities. 

There is a vast amount of literature dealing with alterations of the light forces exerted on atoms due to these features (e.g.\ \cite{Metcalf}). In optical lattices for instance, the interference structure between intersecting monochromatic laser beams is exploited to form a periodic and microscopic trapping potential with large field gradients. In the present experiment the potential depth of each lattice site would be much too shallow to trap atoms at the observed temperatures and the gradient forces average to zero on a length scale larger than the optical wavelength. However, introducing a second frequency to the laser field can drastically change the situation resulting in bichromatic forces (for a recent example see \cite{Feng2017}). For our system with one laser frequency very close to the atomic resonance (i.e.\ the cooler frequency) and the second one shifted off the resonance (the repumper frequency), the bichromaticity results in a rectification of the dipole force, an effect first proposed by \citet{Kazantsev89}. In a simple picture \cite{Grimm90}, this effect can be understood as follows: Depending on the difference of the two wavelengths and the distance to the retro-reflecting mirror, the fields of the two frequencies will feature interference patterns that are offset to one another. Due to the AC-Stark shift the field being off resonance will slightly alter the atomic resonance frequency thereby spatially modulating the effective detuning of the field that is close to resonance and the corresponding dipole force. This way the dipole force, which in a monochromatic standing wave averages to zero on a length scale larger than the wavelength, can form potential wells of much larger size than in a monochromatic optical lattice.

The rectified dipole force has theoretically been proposed earlier enabling the formation of an optical superlattice \cite{Wasik97} which can be used as atom trap similar to a magneto-optical trap. However, the specific considerations in \cite{Wasik97} are quite different to the present scenario mainly regarding the detuning of the off-resonance frequency but also regarding the beam geometries and polarization. A first estimate assuming a one-dimensional standing wave of linear polarization containing the cooler and repumper frequencies shows that the formed potential well can be as deep as 10\,K on a length scale of several decimeters. The spatial size of the potential well seems to be too large to confine the atom cloud to a small volume of few mm in diameter. However, in the present case the field  polarization and intensity distributions result in a mixture of traveling and standing waves, forming a complex three-dimensional field. This makes the implementation of the rectified dipole force in the above classical molasses model extremely challenging. It can be speculated that the atomic dynamics in this field and the balance in the trap is then a result of the interplay of the ``stirring'' due to the spontaneous force and the trapping due to the rectified dipole forces. On the basis of the present analysis, we do not make a conclusive statement here and a critical influence of other unwanted but still present effects (e.g.\ concentric diffraction patterns of the beams due to imperfect beam collimation with spherical lenses) cannot be ruled out.



\section{Conclusion}

In conclusion, we reported on an all-optical near-resonant \lisix\ atom trap that -- in contrast to conventional magneto-optical traps -- does not require magnetic field gradients in the trapping region. This feature along with atom temperatures of only few mK and number densities of about 10$^9$\,cm$^{-3}$ make the present atom trap ideally suited for kinematically complete ion-electron coincidence experiments  in COLTRIM spectrometers. The new trap uses the same hardware as magneto-optical traps and can be realized with only small modifications of laser beam geometries and polarization. Therefore, it can easily be implemented in other existing MOTRIMS experiments.

Although a complete theoretical model of the observed trapping mechanism is still pending, it is evident that the spontaneous force alone (i.e.\ the resonant scattering of photons) is not sufficient to describe the observed features. We believe that the atoms are trapped in a superlattice \cite{Wasik97} formed due to the rectified optical dipole force in the bichromatic light field of the cooling lasers. The fact that similar trapping schemes with other species than lithium have previously been realized makes us confident that the present technique can be exploited for the preparation of a large variety of atomic targets for collision experiments.

\section{Acknowledgments}
\begin{acknowledgments}
This work was supported by the National Science Foundation under grant No.\ 1554776 and by the University of Missouri Research Board.

\end{acknowledgments}

%

\end{document}